\documentclass[prd,showpacs,superscriptaddress,nofootinbib,floatfix,10pt]{revtex4-2} 
\usepackage[utf8]{inputenc}
\usepackage{color}
\usepackage{amsfonts,amsmath,amssymb}
\usepackage{longtable,booktabs}
\usepackage{graphicx}
\usepackage{bm}
\usepackage[colorlinks=true,linkcolor=blue,citecolor=blue]{hyperref}
\usepackage{geometry}
\usepackage{epstopdf}
\usepackage{mathtools}
\usepackage{subfigure}
\usepackage{multirow}
\usepackage{diagbox}
\usepackage{afterpage}
\usepackage{placeins}
\usepackage{float}
\usepackage{mathrsfs}
\usepackage{setspace}
\newcommand{\itp}{\affiliation{CAS Key Laboratory of Theoretical Physics, Institute of Theoretical Physics,\\ Chinese Academy of Sciences, Beijing 100190, China}}

\newcommand{\ucas}{\affiliation{School of Physical Sciences, University of Chinese Academy of Sciences, Beijing 100049, China}}

\newcommand{\qfnu}{\affiliation{College of Physics and Engineering, Qufu Normal University, Qufu 273165, China}}

\newcommand{\qfnucyber}{\affiliation{School of Cyber Science and Engineering, Qufu Normal University, Qufu 273165, China}}

\begin{document}

\title{ Hunting for \texorpdfstring{$X_b$}{} via hidden bottomonium decays \texorpdfstring{$X_b\to \pi\pi\chi_{bJ}$}{}}

\author{Zhao-Sai Jia}\email{jzsqfphys@163.com}\qfnu \itp
\author{Zhen-Hua Zhang}\email{zhangzhenhua@itp.ac.cn} \itp \ucas
\author{Wen-Hua Qin}\email{qwh@qfnu.edu.cn} \qfnucyber 
\author{Gang Li}\email{gli@qfnu.edu.cn} \qfnu 

\begin{abstract}

In this work, we investigate the isospin breaking decay $X_b \to \pi^0\chi_{bJ}$ and the isospin conserved decay $X_b \to \pi\pi\chi_{bJ}$, where $X_b$ is
taken to be the heavy quark flavor symmetry counterpart of $X(3872)$ in the bottomonium
sector as a $B^*\bar B$ molecule candidate. Since the mass of this state may be far below the $B {\bar B}^*$ threshold and the mass difference between
the neutral and charged bottom meson is small compared to the binding energy of the $X_b$, the
isospin-violating decay channel $X_b \to \pi^0 \chi_{bJ}$ would be
highly suppressed. The calculated partial width of $X_b \to \pi\pi\chi_{b1}$ is found to be about tens of $\rm{keVs}$, $1\sim 2$ order(s) of magnitude larger than those of $X_b \to \pi\pi\chi_{b2}$ and $X_b \to \pi\pi\chi_{b0}$.
Taking into account the fact that the total width of
$X_b$ may be smaller than a few MeV like $X(3872)$, the
calculated branching ratios $X_b\to \pi\pi \chi_{b1}$ may reach to orders of $10^{-2}$, which makes it a possible channel for
the experimental searching of the $X_b$.

\end{abstract}

\date{\today}

\maketitle

\section{Introduction}
In the past decades, a large number of so-called $XYZ$ states has been discovered on the experiments, and tremendous effort has been taken to unravel their nature beyond the conventional quark model~\cite{Brambilla:2010cs, Swanson:2006st, Eichten:2007qx, Voloshin:2007dx, Brambilla:2019esw, Godfrey:2008nc, Guo:2017jvc}. In 2003, a narrow structure $X(3872)$ ($\chi_{c1}(3872)$) was reported in the $J/\psi\pi^+\pi^-$ invariant mass distribution in $B^+ \to K^+ J/\psi\pi^+\pi^-$ process by the Belle Collaboration~\cite{Belle:2003nnu}. Subsequently, it was gradually confirmed in the $e^+e^-$ collisions by BaBar Collaboration~\cite{BaBar:2004oro}, and in the $pp/p\bar{p}$ collisions by D0~\cite{D0:2004zmu}, CDF~\cite{CDF:2009nxk}, and LHCb Collaborations~\cite{CMS:2013fpt, LHCb:2013kgk}. Its quantum numbers were pinned down to $J^{PC}=1^{++}$~\cite{LHCb:2015jfc}. The $X(3872)$ may be the most renowned exotic candidates with two salient features. One is its very narrow decay width compared to the typical hadronic width, $\Gamma[X(3872)] < 1.2~\rm{MeV}$; the other is its mass is close to the $D\bar{D}^*$ threshold~\cite{ParticleDataGroup:2022pth}, $M_{X(3872)}-M_{D^0}-M_{D^{*0}}=(-0.12 \pm 0.24)~\rm{MeV}$. Based on these two features, it is naturally suggest that the $X(3872)$ might be a $D\bar{D}^*$ hadronic molecule~\cite{Tornqvist:2003na, Hanhart:2007yq} .

The observation of the $X(3872)$~\cite{Belle:2003nnu} shows that the meson spectroscopy is far more complicated than the naive expectation of the quark model. It is natural to search for the posited bottomonium counterpart of the $X(3872)$ with $J^{PC}=1^{++}$ (called $X_b$ hereafter)~\cite{Ebert:2005nc, Hou:2006it}. As the heavy quark flavor symmetry (HQFS) partner of the $X(3872)$, $X_b$ should share some universal properties with the $X(3872)$. The search for $X_b$ can provide us with important information on the discrimination between the compact multiquark configuration and the loosely bound hadronic molecule configuration for the $X(3872)$~\cite{Belle:2014sys}. The existence of the $X_b$ is predicted both in the molecular interpretation~\cite{Tornqvist:1993ng, Karliner:2015ina, Guo:2013sya, Mutuk:2018zxs} and the tetraquark model~\cite{Ali:2009pi}, with a mass coincide with the $B\bar{B}^*$ threshold~\cite{Tornqvist:1993ng, Guo:2013sya, Karliner:2013dqa, Yamaguchi:2019vea, AlFiky:2005jd, Swanson:2006st, Li:2014uia, Ortega:2021zgk, Guo:2014sca, Zhou:2018hlv, Karliner:2015ina}, or in the $10$ to $11~\rm{GeV/c^2}$ range~\cite{Ebert:2005nc, Ebert:2008se, Ali:2009pi, Matheus:2006xi}, respectively. Such a heavy mass of $X_b$ makes it less likely to be discovered in current electron-positron collision facilities, although the Super KEKB may provide an opportunity for searching it in the radiative decays of the $\Upsilon(5S, 6S)$~\cite{Aushev:2010bq}. The production of the $X_b$ at hadron colliders such as LHCb and Tevatron~\cite{Guo:2014sca, Guo:2013ufa} have been extensively investigated~\cite{Bignamini:2009sk, Esposito:2013ada, Artoisenet:2009wk, Artoisenet:2010uu, Ali:2011qi, Ali:2013xba} and shows sizeable production rate. Therefore, a suitable decay mode from which the $X_b$ state can be reconstructed is imperatively called for. Unlike the $X(3872)$, the isospin breaking decay of the $X_b$ to $\pi^+\pi^-\Upsilon(1S)$ should be highly suppressed since the $X_b$ may be far below the $B\bar{B}^*$ threshold and the mass difference between neutral and charged bottom mesons is very small, which may explain the null result reported by the CMS, ATLAS Collaborations~\cite{CMS:2013ygz, ATLAS:2014mka} for searching the $X_b$ in the $\pi^+\pi^-\Upsilon(1S)$ final state. The isospin conserved hidden bottomonium decay $X_b \to \omega\Upsilon(1S)$ has been investigated in Ref.~\cite{Li:2015uwa} with the intermediate meson-loop (IML) contributions, but no significant signal is observed in the experiments~\cite{Belle:2014sys, Belle-II:2022xdi}. The partial widths of the radioactive decays $X_b\to \gamma \Upsilon(nS)$ was calculated in Ref.~\cite{Li:2014uia} and their magnitude is about $1$~keV. In this work, we will focus on the isospin conserved decay of $X_b \to \pi\pi\chi_{bJ}$ ($J=0, 1, 2$), which could have sizeable branching fractions and be possible channels for the $X_b$ reconstruction.

The $X_b \to \pi\pi\chi_{bJ}$ decays are studied in the bottom IML mechanism. The impact of the IML on the heavy quarkonium transitions has been investigated in the $\Upsilon$ decay processes $\Upsilon(4S) \to h_b(1P, 2P)\pi^+\pi^-$, $\Upsilon(4S) \to \Upsilon(1S, 2S)\pi^+\pi^-$, and $\Upsilon(2S, 3S, 4S) \to \Upsilon(1S, 2S)\pi\pi$~\cite{Chen:2019gfp, Chen:2019gty, Chen:2016mjn}, and gave results consistent with the experimental data. In this work, the partial decay widths of $X_b \to \pi^0\pi^0\chi_{bJ}(1P)$ and $X_b \to \pi^+\pi^+\chi_{bJ}(1P)~(J=0, 1, 2)$ are calculated by using the heavy hadron chiral perturbation theory ($\text{HH}\chi\text{PT}$), including the contribution from the box IML.

The rest of this paper is organized as follows. In Sec.~\ref{sec:Effective Lagrangians and Power Counting}, we introduce the effective Lagrangians and Feynman diagrams for the hadronic decays of the $X_b$ to $\pi^0\chi_{bJ}$ and $\pi\pi\chi_{bJ}$. Our numerical results are presented in Sec.~\ref{sec:Numerical Results}, and a brief summary is given in Sec.~\ref{sec:Summary}.

\section{Effective Lagrangians and Power Counting}\label{sec:Effective Lagrangians and Power Counting}

In this section, we will give the effective Lagrangians and Feynman diagrams for the hadronic decays of $X_b$ to $\pi^0\chi_{bJ}$ and $\pi\pi\chi_{bJ}$. The $X_b$ is assumed to be an $S$-wave molecular state with $J^{PC}=1^{++}$ given by the superposition of $B^0\bar{B}^{*0}+\mathrm{c.c}$ and $B^-B^{*+}+\mathrm{c.c}$ hadronic configurations as~\cite{Li:2015uwa}
\begin{align}
    |X_b\rangle = \frac{1}{\sqrt{2}} \cos{\varphi}(|B^0\bar{B}^{*0}\rangle+|B^{*0}\bar{B}^0\rangle)+\frac{1}{\sqrt{2}} \sin{\varphi}(|B^+B^{*-}\rangle+|B^-B^{*+}\rangle),
\label{eq:X_b}
\end{align}
where the charge conjugation conventions $B^* \overset{C}{\rightarrow} \bar{B}^*$ and $B \overset{C}{\rightarrow} \bar{B}$ is used, and $\varphi$ is a phase angle describing the proportion of neutral and charged constituents, which will be settled in Sec.~\ref{sec:Numerical Results} by the cancellation of the neutral and charged meson loops in the isospin-violating processes $X_b\to \pi^0\chi_{bJ}$. The effective Lagrangian for the $X_b$ coupling to the $B^* \bar{B}$ can be written as 
\begin{align}
\mathcal{L}_{XBB^*}=\frac{g_n}{\sqrt{2}}X^{i \dagger}(B^{*0i}\bar{B}^0 +B^0\bar{B}^{*0i}) + \frac{g_c}{\sqrt{2}}X^{i \dagger}(B^{*+i}B^- +B^+B^{*-i})+\text{H.c.},
\label{eq:L_XBB}
\end{align}
where $g_n$ and $g_c$ denote the coupling constant of the $X_b$ to the neutral and charged bottom meson pairs, respectively. As an isoscalar $B^{\ast}\bar{B}$ molecular state, the $X_b$ state appears as a pole on the real axis in the complex energy plane of the $B^{\ast+}B^{-}-B^{\ast 0}\bar{B}^{ 0}$ coupled channel $T$ matrix with $C=+$ , and the effective couplings $g_n$ and $g_c$ can be derived from the residues of the $T$ matrix elements at the $X_b$ pole and read~\cite{Mehen:2015efa, Meng:2021jnw}
\begin{align}
    g_n = &\, \frac{4\sqrt{\pi \gamma_n}}{\mu_n} \cos{\varphi}, \label{eq:gn}\\   
    g_c = &\, \frac{4\sqrt{\pi \gamma_c}}{\mu_c} \sin{\varphi},
    \label{eq:gc}
\end{align}
where $\gamma_n=\sqrt{2 \mu_n E_{X_b}^n}$ and $\gamma_c=\sqrt{2 \mu_c E_{X_b}^c}$ with $E_{X_b}^n=m_{B^{*0}}+m_{B^0}-m_{X_b}$, $E_{X_b}^c=m_{B^{*+}}+m_{B^+}-m_{X_b}$ the binding energies of the $X_b$ relative to the $B^0\bar{B}^{*0}$ and $B^+B^{*-}$ thresholds, respectively, and $\mu_n=m_{B^{*0}}m_{B^0}/(m_{B^{*0}}+m_{B^0})$, $\mu_c=m_{B^{*+}}m_{B^-}/(m_{B^{*+}}+m_{B^-})$ are the  the reduced masses.

The leading-order heavy hadron chiral perturbation theory (HH$\chi$PT) Lagrangian for mesons containing heavy quarks or antiquarks at rest is~\cite{Fleming:2008yn}
\begin{align}
    &\mathcal{L}_{B^{(*)}B^*\phi} =-g \text{Tr}[H_a^\dagger H_b \vec \sigma \cdot \vec A_{ba}]+g \text{Tr}[\bar{H}_a^\dagger \vec \sigma \cdot \vec A_{ab} \bar{H}_b]+\text{Tr}\left[ \text{H}_a^{\dagger} (i \text{D}_0)_{ba} \text{H}_b \right] +\text{Tr}\left[ \bar{\text{H}}_a^{\dagger} (i \text{D}_0)_{ba} \bar{\text{H}}_b \right],
    \label{eq:L_BBphi}
\end{align}
where $\vec \sigma$ denote the Pauli matrices and $a$ is the light flavor index, the bottom mesons are given by the two-component notation~\cite{Hu:2005gf} as $H_a = \vec V_a \cdot \vec \sigma + P_a$ with $P=(B^-, \bar{B}^0)$ and $V=(B^{*-}, \bar{B}^{*0})$ denote the pseudoscalar and vector heavy mesons, respectively, and the field for the antimesons is $\bar{H}_a=-\vec{\bar{V}}_a \cdot \vec{\sigma} +\bar{P}_a$ with $\bar{P}=(B^+, B^0)$ and $\bar{V}=(B^{*+}, B^{*0})$. The first two terms in Eq.~\eqref{eq:L_BBphi} give the interaction between the heavy mesons and pions. The field $\vec{A}_{ab}=-\vec \bigtriangledown{\Phi}_{ab}/f_{\pi}+\cdots$ is the axial current in the chiral perturbation theory ($\chi$PT) and couples to the heave mesons with the axial coupling $g=0.54$~\cite{Mehen:2015efa}, where $f_{\pi}=130~\rm{MeV}$ is the pion decay constant, and the $\Phi$ field contains the Goldstone bosons as components,
\begin{align}
\Phi=\left( 
\begin{array}{cc}
\frac{1}{\sqrt{2}}\pi^{0} & \pi^{+} \\
\pi^{-} & -\frac{1}{\sqrt{2}}\pi^{0}
\end{array}\right).
\end{align} 
The last two terms in Eq.~\eqref{eq:L_BBphi} describe the $ B^{(*)} B^{(*)}\pi\pi$ interaction, where $ D_0=\partial_0 -V_0 $, $ V_{\mu}= \frac{1}{2}(u^{\dagger} \partial_{\mu}u+ u\partial_{\mu}u^{\dagger})$ is the vector current in the $\chi$PT with $u=\text{exp}(i \Phi/f_{\pi})$.

The Lagrangian couples the $\chi_{bJ}$ to heavy mesons reads
\begin{align}
& \mathcal{L}_{\chi B^{(*)}B^{(*)}}= i \frac{g_1}{2} \text{Tr} \left[ \chi^{i \dagger} H_a \sigma^i \bar{H}_a \right] +\frac{c_1}{2} \text{Tr}\left[ \chi^{i \dagger} H_a \sigma^j \bar{H}_b \right] \varepsilon _{ijk} A_{ab}^k + \text{H.c.},
\label{eq.L_chiBB}
\end{align}
where the $\chi_{bJ}$ field is expressed as~\cite{Fleming:2008yn}
\begin{align}
    \chi^i=\sigma^j \chi^{ij}=\sigma^j \left( \chi_{b2}^{ij} +\frac{1}{\sqrt{2}} \varepsilon ^{ijk} \chi_{b1}^k +\frac{\delta^{ij}}{\sqrt{3}} \chi_{b0} \right),
\end{align}
with the coupling constant $g_1=0.53_{-0.13}^{+0.19}~\rm{GeV}^{-1/2}$~\cite{Chen:2019gfp}.

\begin{figure}[htbp]
      \subfigure[] {
        \includegraphics[scale=0.5]{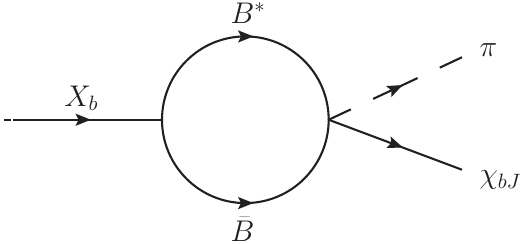}
        \label{fig_2PchibJpi}
    }
     \subfigure[] {
        \includegraphics[scale=0.6]{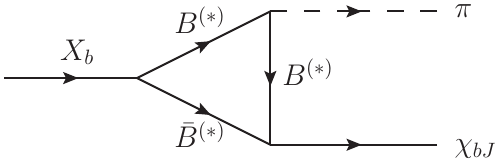}
        \label{fig_3PchibJpi}
    }
    \vspace{0.5 cm}
    
    \subfigure[] {
        \includegraphics[scale=0.55]{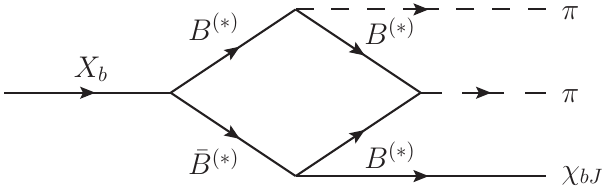}
        \label{fig_4PchibJpipi_b}
    }
    \quad
    \subfigure[] {
        \includegraphics[scale=0.55]{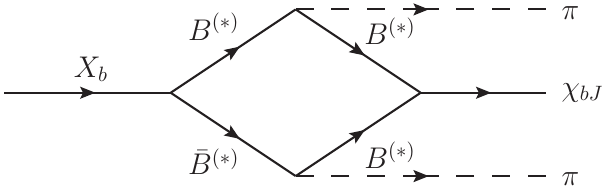}
        \label{fig_4PchibJpipi_c}
    }
    \vspace{0.3 cm}
    
    \subfigure[] {
        \includegraphics[scale=0.6]{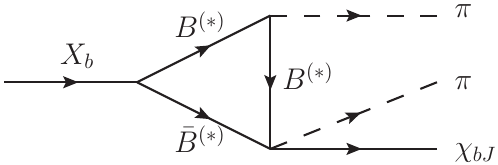}
        \label{fig_3PchibJpipi_b}
    }
    \quad
    \subfigure[] {
        \includegraphics[scale=0.6]{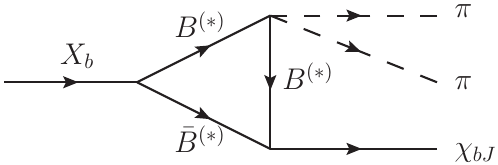}
        \label{fig_3PchibJpipi_t}
    }
    \caption{Feynman diagrams for calculating the partial decay width of $X_b \to \pi^0\chi_{bJ}$ and $ \pi \pi \chi_{bJ}$.}
    \label{fig_XbchibJ}
\end{figure}

The Feynman diagrams of $X_b\to \pi^0\chi_{bJ}$ and $X_b\to \pi\pi\chi_{bJ}$ are shown in Fig.~\ref{fig_XbchibJ} with all the possible combination of the intermediate particles in the loops of each diagram listed in Table~\ref{Tab:loop particle}. In Table~\ref{Tab:loop particle}, the first particle in each square bracket denotes the top (in the two-point bubble), or top left (in the triangle and box) intermediate bottom meson in the corresponding diagram, and the other intermediate bottom mesons in the same diagram are listed in the square bracket in counterclockwise order along the loop.

\begin{table}[htbp]
    \renewcommand\arraystretch{2}
    \centering
    \caption{All the possible combinations of the intermediate heavy mesons for the diagrams in Fig.~\ref{fig_XbchibJ}. The first particle in each square bracket denotes the top (in the two-point bubble), or top left (in the triangle and box) intermediate bottom meson in the corresponding diagram, and the other intermediate bottom mesons in the same diagram are listed in the square bracket in counterclockwise order along the loop.}
    \begin{tabular}{cc}
    \hline\hline
    Fig.1(a) &$[B^*, \bar{B}]$, $[\bar{B}^*, B]$ \\
    \hline
    Fig.1(b) &$[B^*, \bar{B}, B]$, $[\bar{B}^*, B, \bar{B}]$, $[\bar{B}, B^*, \bar{B}^*]$, $[B, \bar{B}^*,B^*]$, $[B^*, \bar{B}, B^*]$, $[\bar{B}^*, B, \bar{B}^*]$ \\
    \hline
    \multirow{3}{*}{Figs.1(c)(d)} &$[B^*, \bar{B}, B^*, B]$, $[B^*, \bar{B}, B, B^*]$, $[B^*, \bar{B}, \bar{B}^*, B^*]$, $[B^*, \bar{B}, B^*, B^*]$, $[B^*, \bar{B}, \bar{B}^*, B]$, $[\bar{B}^*, B, B^*, \bar{B}]$\\
    &$[\bar{B}^*, B, \bar{B}^*, \bar{B}]$, $[\bar{B}^*, B, \bar{B}^*, \bar{B}^*]$, $[\bar{B}^*, B, \bar{B}, \bar{B}^*]$, $[\bar{B}^*, B, B^*, \bar{B}^*]$, $[B, \bar{B}^*, B^*, B^*]$, $[B, \bar{B}^*, \bar{B}^*, B^*]$\\
    &$[B, \bar{B}^*, \bar{B}, B^*]$, $[B, \bar{B}^*, B, B^*]$, $[\bar{B}, B^*, \bar{B}, \bar{B}^*]$, $[\bar{B}, B^*, \bar{B}^*, \bar{B}^*]$, $[\bar{B}, B^*, B^*, \bar{B}^*]$, $[\bar{B}, B^*, B, \bar{B}^*]$\\
    \hline
    Figs.1(e)(f) &$[B^*, \bar{B}, B^*]$, $[B^*, \bar{B}, B]$, $[\bar{B}^*, B, \bar{B}^*]$, $[\bar{B}^*, B, \bar{B}]$, $[B, \bar{B}^*, B^*]$, $[B, \bar{B}^*, B]$, $[\bar{B}, B^*, \bar{B}^*]$, $[\bar{B}, B^*, \bar{B}]$ \\
    \hline\hline
    \end{tabular}
    \label{Tab:loop particle}
\end{table}
Since the $X_b$ is close to the $B\bar{B}^*$ threshold, the velocity of the intermediate heavy meson $v_{B^{(*)}}=\sqrt{\vert E_{X_b} \vert /m_{B^{(*)}}}$ should be smaller than 1 (in units of the speed of light) and become a natural small quantity for the power counting of the diagrams in Fig.~\ref{fig_XbchibJ}. The diagrams can be counted in the powers of $v$ and $p_{\pi}$, where $p_{\pi} \simeq (m_{X_b}-m_{\chi_{bJ}})/2$ is the momentum of the external pion, and $v=(v_{X_b}+v_{\chi_{bJ}})/2$ with $v_{X_b} \simeq 0.03$ and $v_{\chi_{bJ}} \simeq 0.40$ derived from taking the $X_b$ binding energy to be $E_{Xb}^n=5~\mathrm{MeV}$. In each diagram, the nonrelativistic energy counts as $v^2$, each loop integral is at $v^5$, and each nonrelativistic propagator contributes at $v^{-2}$. For the vertices, the $B^{(*)}B^{(*)}\pi\pi$ vertex is proportional to the square of the energy of the pion $E_{\pi}^2 \sim p_{\pi}^2$, the $P$-wave $B^*\bar{B}\pi\chi_{bJ}$ and $B^{(*)}B^{(*)}\pi$ vertices are proportional to $p_{\pi}$, while the $X_b B^{(*)}\bar{B}^{(*)}$ and $B^{(*)}\bar{B}^{(*)}\chi_{bJ}$ vertices are in $S$-wave and count at $v^0p_{\pi}^0$. The diagrams of $X_b \to \pi^0\chi_{bJ}$ shown in Figs.~\ref{fig_2PchibJpi} and \ref{fig_3PchibJpi} scales as $v^5p_{\pi}/(v^2)^2=p_{\pi} v$ and $v^5p_{\pi}/(v^2)^3=p_{\pi}/v$, respectively. For the decay $X_{b}\to \pi\pi\chi_{bJ}$, the box diagrams Figs.~\ref{fig_4PchibJpipi_b} and \ref{fig_4PchibJpipi_c} scales as $v^5 p_{\pi}^2/(v^2)^4=p_{\pi}^2/v^3$, and the triangle diagrams in Figs.~\ref{fig_3PchibJpipi_b} and \ref{fig_3PchibJpipi_t} scale as $v^5p_{\pi}^2/(v^2)^3=p_{\pi}^2/v$ and $v^5p_{\pi}^2/(v^2)^3=p_{\pi}^2/v$, respectively. One can see the contributions from the two point bubble diagrams to the $X_b\to\pi\chi_{bJ}$ decays are suppressed by $v^2$ relative to the triangle diagrams, and the contributions from the triangle diagrams to the $X_b\to\pi\pi\chi_{bJ}$ decays are suppressed by $v^2$ comparing with the box diagrams. Thus, for a rough estimate of the partial decay width, we only consider the contribution of the triangle diagrams to the decay widths of $X_b\to \pi \chi_{bJ}$ and the box diagrams to the decay widths of $X_b\to \pi \pi \chi_{bJ}$ in this work,
and a evaluation of the omitted contributions from the bubble diagrams to the decay widths of $X_b\to \pi \chi_{bJ}$ and the  triangle diagrams to the decay widths of $X_b\to \pi \pi \chi_{bJ}$ is given in Appendix~\ref{sec:two and three point diagram estimate}.

Based on the Lagrangians given above, the loop transition amplitudes in Fig.~\ref{fig_XbchibJ} can be expressed in a general form as follows,
\begin{align}
 \mathcal{A}[X_b \to \pi \chi_{bJ}]=&\, V_1 V_2 V_3 \times I[m_1,m_2,m_3],\\
 \mathcal{A}[X_b \to \pi \pi \chi_{bJ}]=&\, V_1 V_2 V_3 V_4 \times I[m_1,m_2, m_3, m_4],
\end{align}
where $V_n$ ($n=1, 2, 3, 4$) represent the vertex functions for the initial $X_b$, final $\chi_{bJ}$ and $\pi$, respectively. The expressions of the 3-point and 4-point integral functions $I[m_1, m_2, m_3]$, and $I[m_1,m_2, m_3, m_4]$ are given in the Appendix~\ref{sec:loop integrals}, where $m_i~(i=1, 2, 3, 4)$ represents the mass of $i$th particle of each combination of the intermediate states listed in Table~\ref{Tab:loop particle} and all the combinations in each diagram should be summed to get the final amplitude. Note that these amplitudes must be multiplied by a factor of $\sqrt{m_{X_b} m_{\chi_{bJ}}} m_1 m_2 m_3$, or $\sqrt{m_{X_b} m_{\chi_{bJ}}} m_1 m_2 m_3 m_4$ to account for the nonrelativistic normalization of the HH$\chi$PT. With all the amplitudes, the decay rate for the $X_b$ decay is given by
\begin{align}
d\Gamma= \frac{1}{2SM}\frac{1}{2j+1} \sum_{\text{spins}} \left\vert \mathcal{A}\right\vert^2 d\Phi_n,
\label{Eq.Xb decay rate}
\end{align}
where the symmetry factor $S$ is taken to be 2 in the $X_b \to \pi^0\pi^0\chi_{bJ}$ decays considering the identical $\pi^0\pi^0$ particles in the final states, and to be 1 in the decays $X_b \to \pi^0\chi_{bJ}$ and $X_b \to \pi^+\pi^-\chi_{bJ}$, $j$ is the total spin of the initial particle, and there is a sum over all the polarizations of the final-state particles.
The $d\Phi_n$ ($n=2, 3$) is the two-body phase space or three-body phase space, which can be obtained from Ref.~\cite{Jia:2022qwr,Jia:2023hvc}.

\section{Numerical Results}\label{sec:Numerical Results}
In this section, we give the partial decay widths of the $X_b \to \pi^0\chi_{bJ}$, $\pi^0\pi^0\chi_{bJ}$, and $\pi^+\pi^-\chi_{bJ}$. As the binding energy of the $X_b$ is uncertain, covering the mass range of the $X_b$ predicted by the molecular and tetraquark model in Ref.~\cite{Ali:2009pi, Tornqvist:1993ng, Karliner:2015ina, Guo:2013sya} could be a good approximation for calculating the decay widths and might be applicable. In Ref.~\cite{Ali:2009pi}, the mass of the lowest-lying $1^{++}$ $\bar{b}\bar{q}bq$ tetraquark was predicated to be $10504~\rm{MeV}$. In the molecular interpretation, it was predicted to be $10562~\rm{MeV}$, which is approximately $42~\rm{MeV}$ below the $B\bar{B}^*$ threshold~\cite{Tornqvist:1993ng}, and $(10580^{+9}_{-8})~\rm{MeV}$ with a binding energy of $(24^{+8}_{-9})~\rm{MeV}$~\cite{Guo:2013sya}.  Thus, we performed the calculations up to a binding energy of $100~\rm{MeV}$, and choose several illustrative values of $E_{X_b}^n=(2, 5, 10, 25, 50, 100)~\rm{MeV}$ ($E_{Xb}^c=E_{Xb}^n-1.2~\rm{MeV}$) for discussion. 

\begin{figure}[htbp]
      \subfigure[] {
      \includegraphics[scale=0.36]{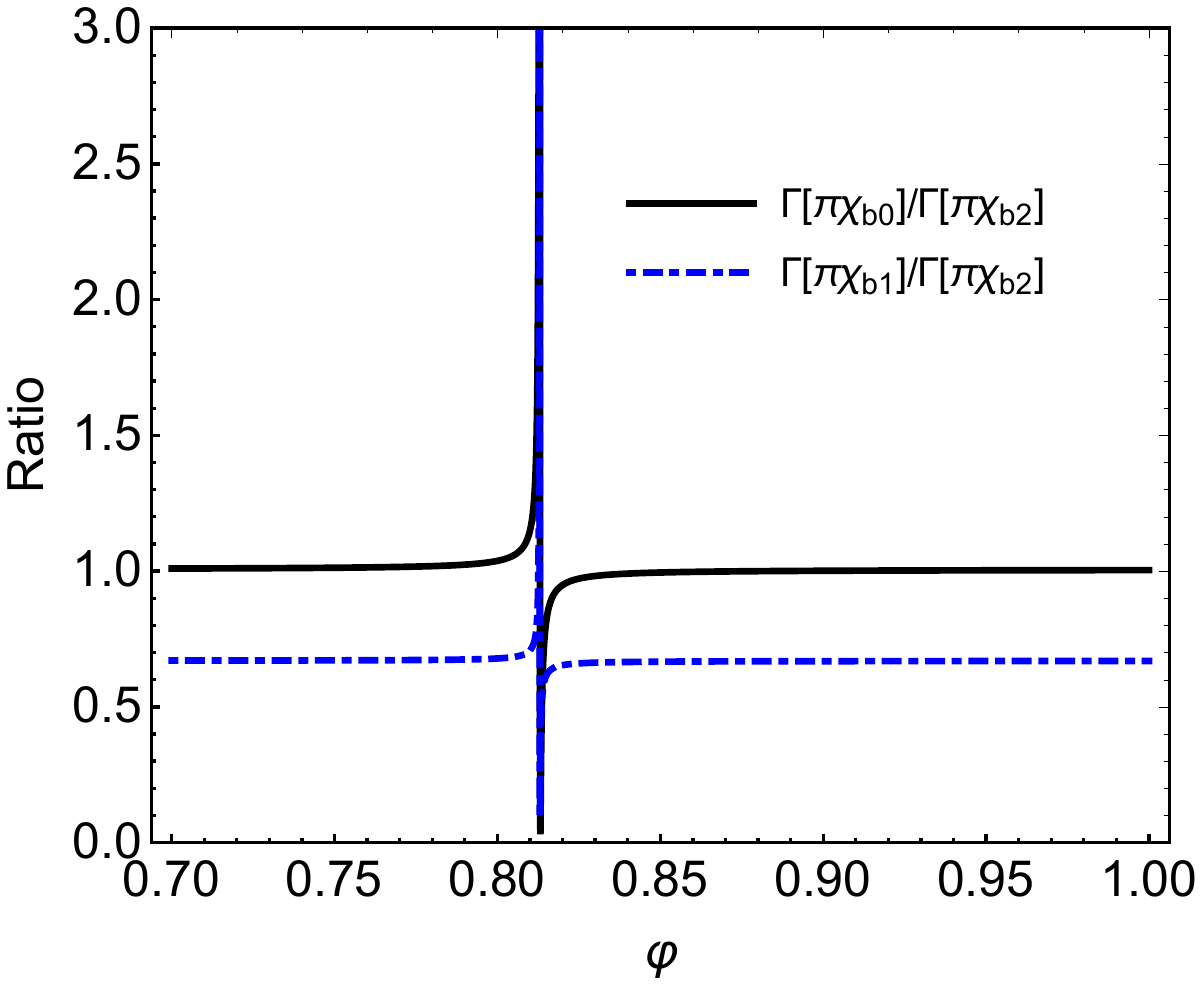}
      \label{fig_XbchibJpi0_Ratio0710}
     } 
      \subfigure[] {
      \includegraphics[scale=0.35]{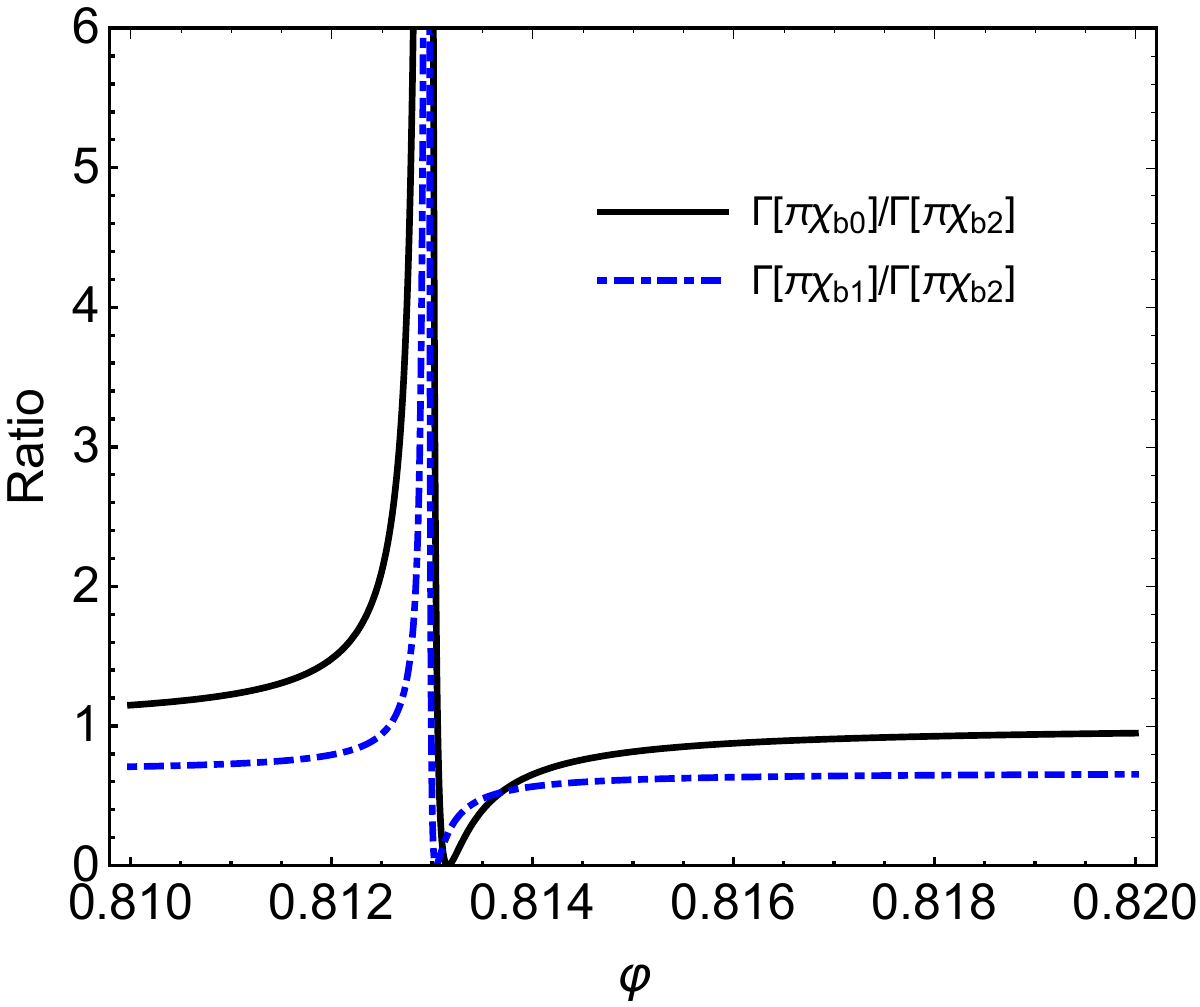}
      \label{fig_XbchibJpi0_Ratio0812}
      }
    \caption{$\Gamma[\pi^0\chi_{b0}]/\Gamma[\pi^0\chi_{b2}]$ and $\Gamma[\pi^0\chi_{b1}]/\Gamma[\pi^0\chi_{b2}]$ as a function of $\varphi$.}
    \label{fig_XbchibJpi0Ratio}
\end{figure}

It is likely that the $X_b$ is below the $B^*\bar B$ threshold and the mass difference between
the neutral and charged bottom meson is small compared to the binding energy of the $X_b$~\cite{Tornqvist:1993ng, Karliner:2015ina, Guo:2013sya,Ali:2009pi}, the isospin violating decay mode $X_b \to \pi^0 \chi_{bJ}$ would be greatly suppressed.  Based on this point, we can determine the charged and neutral bottom meson components of $X_b$ to some extent. If we assume that the proportions of the neutral and charged components are the same. i.e., $\varphi=\pi/4$ in Eq.~(\ref{eq:X_b}), the calculated partial decay widths are
\begin{align}
    \Gamma[\pi^0\chi_{b0}] =&\, 13.10~\rm{keV},\\
    \Gamma[\pi^0\chi_{b1}] =&\, 8.64~\rm{keV},\\
    \Gamma[\pi^0\chi_{b2}] =&\, 12.83~\rm{keV},
\end{align}
which are too large for the isospin breaking processes as the neutral and charged meson loops are not canceled properly. 
The cancellation can be arranged by finely tuning the angle $\varphi$ to adjust the charged and neutral mesons components of $X_b$~\cite{Mehen:2015efa}. The ratios $\Gamma[\pi^0\chi_{b0}]/\Gamma[\pi^0\chi_{b2}]$ and $\Gamma[\pi^0\chi_{b1}]/\Gamma[\pi^0\chi_{b2}]$ with $\varphi\in [0.7,1.0]$ are shown in Fig.~\ref{fig_XbchibJpi0Ratio}.
One has $\Gamma[\pi^0\chi_{b0}]/\Gamma[\pi^0\chi_{b2}] \simeq 1.0$ and $\Gamma[\pi^0\chi_{b1}]/\Gamma[\pi^0\chi_{b2}] \simeq 0.7$ for most values of $\varphi$ except for $\varphi \in[0.81, 0.82$] where the charged and the neutral loops cancel with each other.

\begin{table}[htbp]
    \renewcommand\arraystretch{2}
    \centering
    \caption{Predicted partial widths (in units of $\rm{keV}$) of the $X_b$ decays. The units of the binding energy $E_{X_b}^n$ in first column are $\rm{MeV}$.}
    \begin{tabular}{cccccccccc}
    \hline\hline
     &$\Gamma^{\text{B}}[\pi^0\pi^0\chi_{b0}]$ &$\Gamma^{\text{B}}[\pi^0\pi^0\chi_{b1}]$ &$\Gamma^{\text{B}}[\pi^0\pi^0\chi_{b2}]$
     &$\Gamma^{\text{B}}[\pi^+\pi^-\chi_{b0}]$ &$\Gamma^{\text{B}}[\pi^+\pi^-\chi_{b1}]$ &$\Gamma^{\text{B}}[\pi^+\pi^-\chi_{b2}]$\\
    \hline
    $E_{X_b}^n=2~\rm{MeV}$  &$0.13$ &$10.00$ &$2.07$ &$0.24$ &$18.36$ &$3.89$\\
    $E_{X_b}^n=5~\rm{MeV}$  &$0.18$ &$14.60$ &$3.11$ &$0.33$ &$27.61$ &$5.85$\\
    $E_{X_b}^n=10~\rm{MeV}$ &$0.19$ &$16.98$ &$3.62$ &$0.35$ &$32.09$ &$6.80$\\
    $E_{X_b}^n=25~\rm{MeV}$ &$0.15$ &$16.52$ &$3.50$ &$0.27$ &$32.16$ &$6.57$\\
    $E_{X_b}^n=50~\rm{MeV}$ &$0.09$ &$12.54$ &$2.54$ &$0.15$ &$23.51$ &$4.74$\\
    $E_{X_b}^n=100~\rm{MeV}$&$0.02$ &$5.62$  &$1.10$ &$0.04$ &$10.41$ &$2.03$\\  
    \hline\hline
    \end{tabular}
    \label{Tab:Xb partial widths}
\end{table}

After the careful arrangement of the charged and neutral components in $X_b$, we give the partial widths of the $X_b\to \pi\pi\chi_{bJ}$ process with $\varphi=0.813 \pm 0.005$, which are isospin conserved and not sensitive to the angle $\varphi$ in the neighborhood of $\varphi=0.813$. The partial widths of the $X_b \to \pi\pi\chi_{bJ}$ with $E_{X_b}^n=(2, 5, 10, 25, 50, 100)~\mathrm{MeV}$ are listed in Table~\ref{Tab:Xb partial widths}, where the partial decay widths of $X_b \to \pi\pi\chi_{bJ}~(J=1, 2)$ are about tens of $\rm{keVs}$, while the partial decay widths of $X_b \to \pi\pi\chi_{b0}$ is two orders of magnitude smaller because it is in higher partial waves as analyzed in the follows. The quantum numbers of the identical particle ($\pi^0\pi^0$) system are $I^G(J^{PC})=0^+(L^{++})~(L=\text{even})$. Considering the conservation of the angular momentum and the $P$-parity, in the $X_b \to \pi^0\pi^0\chi_{b0}$ and $X_b \to \pi^0\pi^0\chi_{b2}$ processes, the two pions in the identical particle system should be at least in $D$-wave, and this system should also be at least in $D$-wave and $S$-wave with the $\chi_{b0}$ and $\chi_{b2}$, respectively. While in the $X_b \to \pi^0\pi^0\chi_{b1}$ decay, both the two pions and the two-pion system with the $\chi_{b1}$ can be in $S$-wave, therefore the partial widths of $X_b \to \pi^0\pi^0\chi_{b0}$ are supressed by the higher partial waves comparing with those of $X_b$ decaying into $\pi^0\pi^0\chi_{b1}$ and $\pi^0\pi^0\chi_{b2}$. For the decays $X_b \to \pi^+\pi^-\chi_{bJ}$, the quantum numbers of the ($\pi^+\pi^-$) system satisfy $L+I=\mathrm{even}$ with $I$ the total isospin of the ($\pi^+\pi^-$) system. For $I=0$,
the discussions are same as the decays to $\pi^0\pi^0\chi_{bJ}$.
For $I=1$, the two charged pions can be in $P$-wave and the two-pion system and the $\chi_{b0}$ can also be in $P$-wave, and the partial width of the $X_b \to \pi^+\pi^-\chi_{b0}$ process is still suppressed as it is isospin-violated.  And one can see that $\Gamma[\pi^+\pi^-\chi_{bJ}]/\Gamma[\pi^0\pi^0\chi_{bJ}] \simeq 2$, same as the $X(3872)$ case~\cite{Fleming:2008yn}.

\begin{figure}[htbp]
      \subfigure[] {
      \includegraphics[scale=0.3]{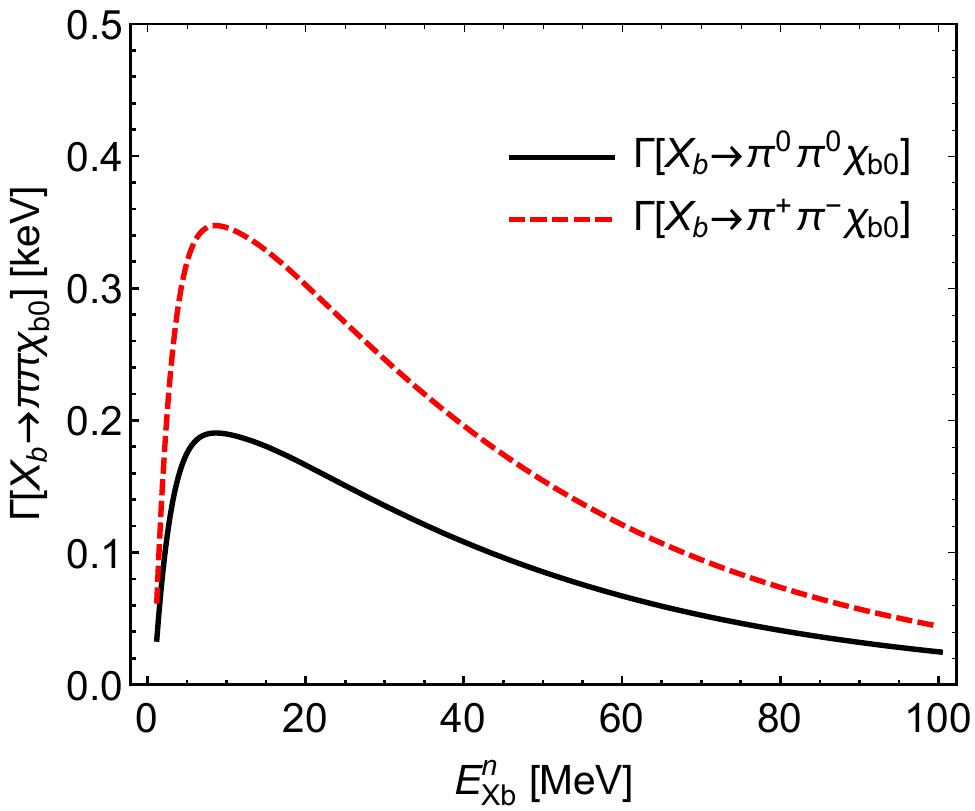}
      \label{fig_Xbchib0pipiwidth}
     } 
      \subfigure[] {
      \includegraphics[scale=0.3]{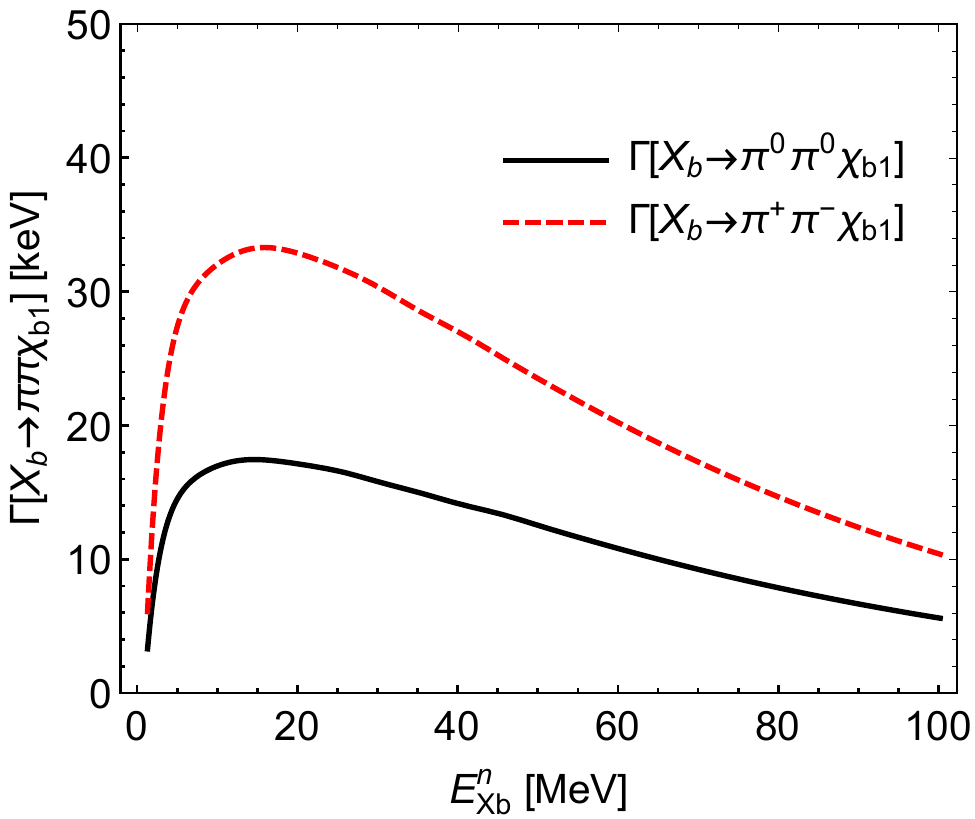}
      \label{fig_Xbchib1pipiwidth}
      }
      \subfigure[] {
      \includegraphics[scale=0.3]{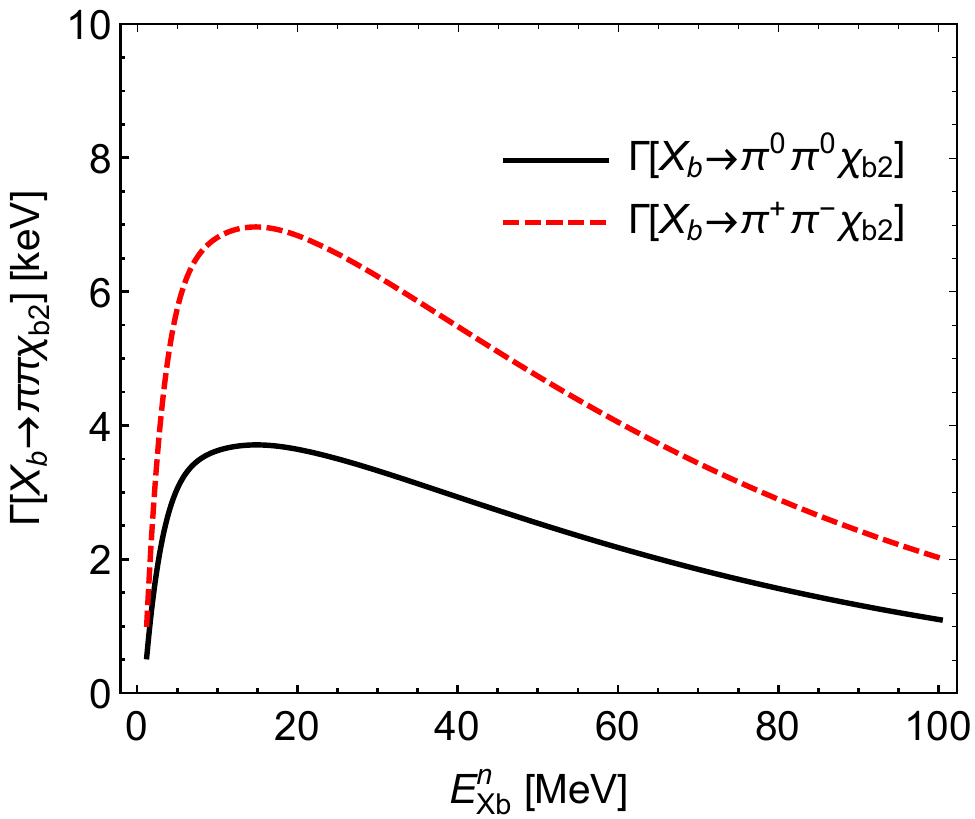}
      \label{fig_Xbchib2pipiwidth}
      }\caption{The partial decay widths of $X_b \to \pi\pi\chi_{bJ}$ as a function of the binding energy $E_{X_b}^n$.}
    \label{fig_XbchibJpipiwidth}
\end{figure}

To see the relation between the partial widths of $X_b \to \pi\pi\chi_{bJ}~(J=0, 1, 2)$ and the binding energy of the $X_b$ more precisely, the partial widths versus the binding energy $E_{X_b}^n\in [2,100]~\mathrm{MeV}$ are demonstrated in Fig.~\ref{fig_XbchibJpipiwidth}, where the partial widths first increase and then decrease with the $E_{X_b}^n$ increases. This is because the binding energy dependence of the partial width can be influenced by the coupling strength of $X_b$ in Eqs.~\eqref{eq:gn},~\eqref{eq:gc}, and the threshold effects. As the binding energy $E_{X_b}^n$ increases, the coupling strength of $X_b$ increases, while the threshold effects decreases. In the small $E_{X_b}^n$ region, the contribution from the the coupling strength of $X_b$ is dominant, while that from the threshold effects plays an important role in the large $E_{X_b}^n$ region.

One can see that unlike the $X_b \to \pi^+\pi^-\Upsilon$ which is suppressed by the isospin symmetry, the partial decay widths of the isospin conserved process $X_b \to \pi\pi\chi_{b1}$ is at the same order of magnitude with the partial width of $X_b \to \Upsilon(1S) \omega$ calculated in Ref.~\cite{Li:2015uwa}, and it could be better channel for the $X_b$ searching as the $\omega$ needs to be reconstructed through the $\pi\pi\pi$ or $\pi^0\gamma$ final states. Therefore our results can be helpful for hunting the $X_b$ in the experiments.

\section{Summary}\label{sec:Summary}

In this work, we investigated the isospin breaking decay $X_b \to \pi^0\chi_{bJ}$ and the isospin conserved decay $X_b \to \pi\pi\chi_{bJ}$, where $X_b$ is
taken to be the HQFS counterpart of $X(3872)$ in the bottomonium
sector as a meson-meson molecule candidate. Since the mass of this state may be far below the $B {\bar B}^*$ threshold, the
isospin-violating decay channel $X_b \to \pi^0 \chi_{bJ}$ would be
highly suppressed and stimulate the importance of the isospin-conserved decay channel $X_b \to \pi\pi\chi_{bJ}$. The isospin-violating decay channel $X_b \to \pi^0 \chi_{bJ}$ can helps us determine the charged and neutral components of $X_b$ to some extent.
For the isospin conserved processes,
the calculated partial widths of $X_b \to \pi\pi\chi_{bJ}~(J=0,1,2)$ are about less than 1 keV, tens of $\rm{keVs}$, and a few $\rm{keVs}$, respectively. 
The partial width of $X_b \to \pi\pi\chi_{b1}$ is found to be about tens of $\rm{keVs}$, $1\sim 2$ order(s) of magnitude larger than those of $X_b \to \pi\pi\chi_{b2}$ and $X_b \to \pi\pi\chi_{b0}$.
Taking into account the fact that the total width of
$X_b$ may be smaller than a few MeV like $X(3872)$, the
calculated branching ratios $X_b\to \pi\pi \chi_{b1}$ may reach to orders of $10^{-2}$, which makes it a possible channel for
the experimental searching of the $X_b$. 
These studies may help us investigate the $X_b$ deeply.
The experimental observation of $X_b$ will provide us further
insight into the spectroscopy of exotic states and is helpful to probe the structure of the states connected by the heavy
quark symmetry.

\section{ACKNOWLEDGMENTS}\label{sec: ACKNOWLEDGMENTS}

The authors thank Yun-Hua Chen, Qi Wu, and Shi-Dong Liu for useful discussions, and thank the anonymous referee for useful comments and suggestions. This work is partly supported by the National Natural Science Foundation of China under Grant Nos.
12075133, 11835015, and 12047503, and by the Natural Science
Foundation of Shandong Province under Grant Nos. ZR2021MA082, and ZR2022ZD26. It is also supported by Taishan
Scholar Project of Shandong Province (Grant No.tsqn202103062),
the Higher Educational Youth Innovation Science and Technology
Program Shandong Province (Grant No. 2020KJJ004).

\appendix
\section{2-POINT, 3-POINT and 4-POINT LOOP INTEGRALS}\label{sec:loop integrals}

In this section, we derive the 2-point, 3-point and 4-point loop integral in the rest frame of the decay particle $(p=(M,0))$. The 2-point loop integral can be written as~\cite{Guo:2010ak}
\begin{align}
I[m_1,m_2]&=i\int \frac{d^4l}{(2 \pi)^4} \frac{1}{[l^2-m_1^2+i\epsilon][(p-l)^2-m_2^2+i\epsilon]}\nonumber \\
&=\frac{i}{4 m_1 m_2}\int \frac{d^4l}{(2 \pi)^4} \frac{1}{[l^0-m_1-\frac{\vec{l}^2}{2 m_1}+i \epsilon][M-l^0-m_2-\frac{\vec{l}^2}{2 m_2}+i \epsilon]}\nonumber \\
&=\frac{1}{4 m_1 m_2}\int \frac{d^{3}l}{(2 \pi)^3} \frac{1}{b_{12}+\frac{\vec{l}^{2}}{2 \mu_{12}}-i \epsilon}\nonumber \\
&=\frac{2 \mu_{12}}{4 m_1 m_2} \int \frac{d^3l}{(2 \pi)^3}\frac{1}{\vec{l}^{2}+c_1-i \epsilon}\nonumber\\
&=\frac{\mu_{12}}{4 \pi^2 m_1 m_2} \left[ \Lambda-\sqrt{c_1-i \epsilon} \tan^{-1}\left(\frac{\Lambda}{\sqrt{c_1-i \epsilon}}\right) \right] \ ,
\label{Eq:2-piont loop_integral}
\end{align}
where $\mu_{ij}=m_{i} m_{j} /\left(m_{i}+m_{j}\right)$ are the reduced masses, $b_{12}=m_1+m_2-M$, and $c_1=2 \mu_{12} b_{12}$. The cutoff $\Lambda$ is taken to be $1~\rm{GeV}$.

The scalar 3-point loop integral is ultraviolet (UV) convergent and can be worked out as~\cite{Guo:2010ak}
\begin{align}
I[m_1,m_2,m_3]&=i\int \frac{d^4l}{(2 \pi)^4} \frac{1}{[l^2-m_1^2+i\epsilon][(p-l)^2-m_2^2+i\epsilon][(l-q)^2-m_3^2+i\epsilon]}\nonumber \\
&=\frac{i}{8 m_1 m_2 m_3}\int \frac{d^4l}{(2 \pi)^4} \frac{1}{[l^0-m_1-\frac{\vec{l}^2}{2 m_1}+i \epsilon][M-l^0-m_2-\frac{\vec{l}^2}{2 m_2}+i \epsilon][l^0-q^0-m_3-\frac{(\vec{l}-\vec{q}\,)^2}{2 m_3}+i \epsilon]}\nonumber \\
&=\frac{1}{8 m_1 m_2 m_3}\int \frac{d^{3}l}{(2 \pi)^3} \frac{1}{\left(b_{12}+\frac{\vec{l}^{2}}{2 \mu_{12}}-i \epsilon\right)\left[b_{23}+\frac{\vec{l}^{2}}{2 m_{2}}+\frac{\left(\vec{l}-\vec{q}\right)^{2}}{2 m_{3}}-i \epsilon\right]}\nonumber \\
&=\frac{4 \mu_{12} \mu_{23}}{8 m_1 m_2 m_3} \int \frac{d^3l}{(2 \pi)^3} \frac{1}{\left(\vec{l}^2+c_1-i \epsilon\right)\left(\vec{l}^2-\frac{2 \mu_{23}}{m_3} \vec{l} \cdot \vec{q}+c_2-i \epsilon\right)}\nonumber\\
&=\frac{4 \mu_{12} \mu_{23}}{8 m_1 m_2 m_3}\int_0^1 d x \int \frac{d^3 l}{(2 \pi)^3} \frac{1}{\left[\vec{l}^2-a x^2+\left(c_2-c_1\right) x+c_1-i \epsilon\right]^2} \nonumber \\
&=\frac{1}{8 m_1 m_2 m_3}\frac{4 \mu_{12} \mu_{23}}{(4 \pi)^{3/ 2}} \Gamma\left(\frac{1}{2}\right) \int_0^1 d x\left[-a x^2+\left(c_2-c_1\right) x+c_1-i \epsilon\right]^{1/ 2} \nonumber \\
&=\frac{\mu_{12} \mu_{23}}{16 \pi m_1 m_2 m_3} \frac{1}{\sqrt{a}}\left[\tan ^{-1}\left(\frac{c_2-c_1}{2 \sqrt{a c_1}}\right)+\tan ^{-1}\left(\frac{2 a+c_1-c_2}{2 \sqrt{a\left(c_2-a\right)}}\right)\right],
\label{Eq:3-piont loop_integral}
\end{align}
where $\mu_{ij}=m_{i} m_{j} /\left(m_{i}+m_{j}\right)$ are the reduced masses, $b_{12}=m_1+m_2-M$, $b_{23}=m_{2}+m_{3}+q^0-M$, and 
\begin{align}
 a=\left(\frac{\mu_{23}}{m_3}\right)^2 \vec{q}^2, \quad c_1=2 \mu_{12} b_{12}, \quad c_2=2 \mu_{23} b_{23}+\frac{\mu_{23}}{m_3} \vec{q}^2.
\end{align}

\begin{figure}[htbp]
    \subfigure[] {
        \includegraphics[scale=0.5]{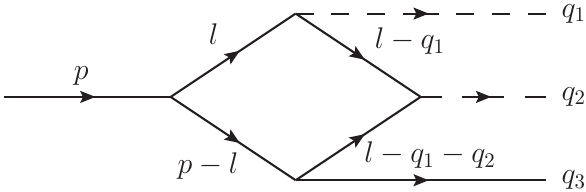}
        \label{fig_4PA1}
    }
    \quad
    \subfigure[] {
        \includegraphics[scale=0.5]{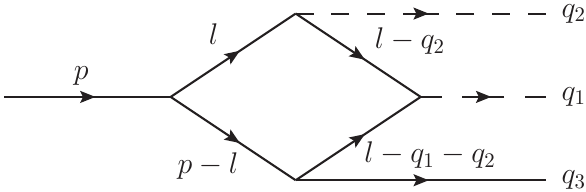}
        \label{fig_4PA1p}
    }
    \quad
    \subfigure[] {
        \includegraphics[scale=0.5]{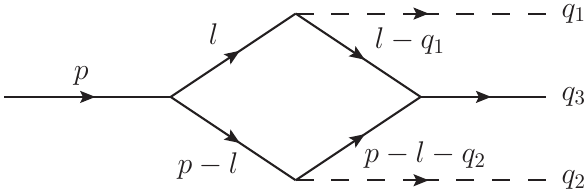}
        \label{fig_4PA2}
    }
    \caption{Kinematics used for calculating 4-point integrals.}
    \label{fig_4PA}
\end{figure}

The four-point integrals in the box diagrams are given in Ref.~\cite{Chen:2019gfp}. For the Fig.~\ref{fig_4PA1}, the initial particle at rest $[p=(M,0)]$ reads
\begin{align}
 & I_1[m_1, m_2, m_3, m_4] \nonumber\\
 &\, \equiv i\int \frac{d^4 l}{(2 \pi)^4} \frac{1}{[l^2-m_1^2+i \epsilon][(p-l)^2-m_2^2+i \epsilon
 ][(l-q_1-q_2)^2-m_3^2+i \epsilon][(l-q_1)^2-m_4^2+i \epsilon]}\nonumber\\
 &\, \simeq \frac{-i}{16 m_1 m_2 m_3 m_4}\nonumber\\
 &\, \times \int \frac{d^4 l}{(2 \pi)^4} \frac{1}{[l^0-\frac{\vec{l}^2}{2 m_1}-m_1+i \epsilon][l^0-M+\frac{\vec{l}^2}{m_2}+m_2-i \epsilon][l^0-q_1^0-q_2^0-\frac{(\vec{l}+\vec{q}_3)^2}{2 m_3}-m_3+i \epsilon]}\nonumber\\
 &\, \times \frac{1}{[l^0-q_1^0-\frac{(\vec{l}-\vec{q}_1)^2}{2 m_4}-m_4+i \epsilon]}\nonumber\\
 &\, =\frac{-\mu_{12} \mu_{23} \mu_{24}}{2 m_1 m_2 m_3 m_4} \int \frac{d^3 l}{(2 \pi)^3}\frac{1}{[\vec{l}^2+c_{12}-i \epsilon][\vec{l}^2+2\frac{\mu_{23}}{m_3}\vec{l} \cdot \vec{q}_3+c_{23}-i \epsilon][\vec{l}^2-2\frac{\mu_{24}}{m_4} \vec{l} \cdot \vec{q}_1 +c_{24}-i\epsilon]},
\end{align}
where
\begin{align}
    c_{12} \equiv &\, 2 \mu_{12} (m_1+m_2-M), \quad c_{23} \equiv 2 \mu_{23} (m_2+m_3-M+q_1^0+q_2^0+\frac{\vec{q}_3^{\,2}}{2 m_3}),\nonumber\\
    c_{24} \equiv &\, 2 \mu_{24} (m_2+m_4-M+q_1^0+\frac{\vec{q}_1^{\,2}}{2 m_4}), \quad \mu_{ij}=\frac{m_i m_j}{m_i+m_j}.
\end{align}
The $m_1$ denotes the mass of the top-left intermediate-bottom mesons, and the mass of the other intermediate bottom mesons are labeled as $m_2$, $m_3$, and $m_4$, in counterclockwise order.

As the crossed diagram of the Fig.~\ref{fig_4PA1p} with $p_1 \leftrightarrow p_2$, the scalar integral of the Fig.~\ref{fig_4PA1p} reads
\begin{align}
 & I_1^{\prime}[m_1, m_2, m_3, m_4] =\frac{-\mu_{12} \mu_{23} \mu_{24}}{2 m_1 m_2 m_3 m_4} \int \frac{d^3 l}{(2 \pi)^3}\frac{1}{[\vec{l}^2+c_{12}-i \epsilon][\vec{l}^2+2\frac{\mu_{23}}{m_3}\vec{l} \cdot \vec{q}_3+c_{23}-i \epsilon][\vec{l}^2-2\frac{\mu_{24}}{m_4} \vec{l} \cdot \vec{q}_2 +c_{24}^{\prime}-i\epsilon]},
\end{align}
where
\begin{align}
    c_{24}^{\prime} \equiv &\, 2 \mu_{24} (m_2+m_4-M+q_2^0+\frac{\vec{q}_2^{\,2}}{2 m_4}).
\end{align}
For the Fig.~\ref{fig_4PA2},
\begin{align}
 & I_2[m_1, m_2, m_3, m_4] \nonumber\\
 &\, \equiv i\int \frac{d^4 l}{(2 \pi)^4} \frac{1}{[l^2-m_1^2+i \epsilon][(p-l)^2-m_2^2+i \epsilon
 ][(p-l-q_2)^2-m_3^2+i \epsilon][(l-q_1)^2-m_4^2+i \epsilon]}\nonumber\\
 &\, \simeq \frac{-i}{16 m_1 m_2 m_3 m_4}\nonumber\\
 &\, \times \int \frac{d^4 l}{(2 \pi)^4} \frac{1}{[l^0-\frac{\vec{l}^2}{2 m_1}-m_1+i \epsilon][l^0-M+\frac{\vec{l}^2}{m_2}+m_2-i \epsilon][l^0+q_2^0-M-\frac{(\vec{l}+\vec{q}_2)^2}{2 m_3}+m_3-i \epsilon]}\nonumber\\
 &\, \times \frac{1}{[l^0-q_1^0-\frac{(\vec{l}-\vec{q}_1)^2}{2 m_4}-m_4+i \epsilon]}\nonumber\\
 &\, =\frac{-\mu_{12} \mu_{34}}{2 m_1 m_2 m_3 m_4} \int \frac{d^3 l}{(2 \pi)^3}\frac{1}{[\vec{l}^2+c_{12}-i \epsilon][\vec{l}^2-\frac{2\mu_{34}}{m_4}\vec{l} \cdot \vec{q}_1+\frac{2\mu_{34}}{m_3}\vec{l} \cdot \vec{q}_2+c_{34}-i \epsilon]}\nonumber\\
 &\, \times \left[ \frac{\mu_{24}}{[\vec{l}^2-\frac{2\mu_{24}}{m_4} \vec{l} \cdot \vec{q}_1 +c_{24}-i\epsilon]}+ \frac{\mu_{13}}{[\vec{l}^2+\frac{2\mu_{13}}{m_3} \vec{l} \cdot \vec{q}_2 +c_{13}-i\epsilon]}\right]    
\end{align}
where
\begin{align}
c_{34} \equiv &\, 2\mu_{34}(m_3+m_4-q_3^0+\frac{\vec{q}_1^{\,2}}{2 m_4}+\frac{\vec{q}_2^{\,2}}{2 m_3}), \quad c_{13} \equiv 2 \mu_{13} (m_1+m_3-M+q_2^0+\frac{\vec{q}_2^{\,2}}{2 m_3}).
\end{align}

\section{Evaluation of the triangle diagram contribution to \texorpdfstring{$X_b \to \pi \pi \chi_{bJ}$}{} and the bubble diagram contribution to \texorpdfstring{$X_b \to \pi \chi_{bJ}$}{}}\label{sec:two and three point diagram estimate}

In this section, we give a rough estimate of the contributions of the omitted triangle diagrams in Figs.~\ref{fig_3PchibJpipi_b} and \ref{fig_3PchibJpipi_t} and the omitted two point bubble diagrams in Fig.~\ref{fig_2PchibJpi} to the decay widths of $X_b \to \pi \pi \chi_{bJ}$ and $X_b \to  \pi \chi_{bJ}$,
respectively.

Unfortunately, the bubble diagram in Fig.~\ref{fig_2PchibJpi} and the triangle diagram in Fig.~\ref{fig_3PchibJpipi_b} contain an unknown coupling $c_1$. Hence, we can only give a quantitative estimate of their contributions based on the power counting and naturalness. In our power counting scheme, the ratio of the contributions from the box diagrams to those from the triangle diagrams is 
\begin{equation}
    1:v^2.
\end{equation}
More specifically, considering the effective couplings $c_1$ and $g_1$, the ratio reads
\begin{equation}
 1:\frac{m_{B^{(\ast)}}c_1}{g_1}v^2,
\end{equation}
where the $m_{B^{(*)}}$ comes from the difference between the triangle and box IML to match the dimensions of $c_1$ and $g_1$.
To be consistent with this power counting, it is natural to take $c_1 \simeq g_1 / m_{B^*}$ to estimate the contributions from the triangle diagrams, and the results are shown in Table~\ref{Tab:Xb partial widths triangle diagrams mBstar}. The $X_b\to \pi\pi\chi_{b1}$ decay also contains the contributions of the triangle diagrams in Fig.~\ref{fig_3PchibJpipi_t}, which is free of the unknown $c_1$. The contributions of Figs.~\ref{fig_3PchibJpipi_b} and~\ref{fig_3PchibJpipi_t} to the decay widths are at the same orders of magnitude in our power counting, which demands $c_1\simeq g_1/(\sqrt{10}m_{B^{\ast}})$. The decay widths with such a coupling are shown in Table~\ref{Tab:Xb partial widths triangle diagrams sqrt ten mBstar}. 
From these evaluations, the contributions of the triangle diagrams to the decay widths are estimated to be $1\sim 3$ magnitude smaller than the box diagram contributions. 
\begin{table}[htbp]
    \renewcommand\arraystretch{2}
    \centering
    \caption{Predicted partial widths (in units of $\rm{keV}$) of the $X_b$ decays with $c_1 = g_1 / m_{B^*}$.}
    \begin{tabular}{cccccccccc}
    \hline\hline
     &$\Gamma^{\text{T}}[\pi^0\pi^0\chi_{b0}]$ &$\Gamma^{\text{T}}[\pi^0\pi^0\chi_{b1}]$ &$\Gamma^{\text{T}}[\pi^0\pi^0\chi_{b2}]$
     &$\Gamma^{\text{T}}[\pi^+\pi^-\chi_{b0}]$ &$\Gamma^{\text{T}}[\pi^+\pi^-\chi_{b1}]$ &$\Gamma^{\text{T}}[\pi^+\pi^-\chi_{b2}]$\\
    \hline
    $E_{X_b}^n=2~\rm{MeV}$&$0.003$ &$0.754$ &$0.408$ &$0.006$ &$2.273$ &$0.770$\\
    $E_{X_b}^n=5~\rm{MeV}$&$0.005$ &$1.214$ &$0.659$ &$0.009$ &$2.296$ &$1.242$\\
    $E_{X_b}^n=10~\rm{MeV}$&$0.005$ &$1.516$ &$0.824$ &$0.009$ &$2.946$ &$1.551$\\
    $E_{X_b}^n=25~\rm{MeV}$&$0.004$ &$1.687$ &$0.915$ &$0.008$ &$3.403$ &$1.784$\\
    $E_{X_b}^n=50~\rm{MeV}$&$0.003$ &$1.433$ &$0.770$ &$0.005$ &$2.938$ &$1.438$\\
    $E_{X_b}^n=100~\rm{MeV}$&$0.001$ &$0.783$ &$0.409$ &$0.002$ &$1.624$ &$0.753$\\ 
    \hline\hline
    \end{tabular}
    \label{Tab:Xb partial widths triangle diagrams mBstar}
\end{table}

\begin{table}[htbp]
    \renewcommand\arraystretch{2}
    \centering
    \caption{Predicted partial widths (in units of $\rm{keV}$) of the $X_b$ decays with $c_1 = g_1 / (\sqrt{10} m_{B^*})$.}
    \begin{tabular}{cccccccccc}
    \hline\hline
     &$\Gamma^{\text{T}}[\pi^0\pi^0\chi_{b0}]$ &$\Gamma^{\text{T}}[\pi^0\pi^0\chi_{b1}]$ &$\Gamma^{\text{T}}[\pi^0\pi^0\chi_{b2}]$
     &$\Gamma^{\text{T}}[\pi^+\pi^-\chi_{b0}]$ &$\Gamma^{\text{T}}[\pi^+\pi^-\chi_{b1}]$ &$\Gamma^{\text{T}}[\pi^+\pi^-\chi_{b2}]$\\
    \hline
    $E_{X_b}^n=2~\rm{MeV}$&$3\times 10^{-4}$ &$0.075$ &$0.041$ &$6\times 10^{-4}$ &$0.990$ &$0.077$\\
    $E_{X_b}^n=5~\rm{MeV}$&$5\times 10^{-4}$ &$0.121$ &$0.066$ &$9\times 10^{-4}$ &$0.230$ &$0.124$\\
    $E_{X_b}^n=10~\rm{MeV}$&$5\times 10^{-4}$ &$0.152$ &$0.082$ &$9\times 10^{-4}$ &$0.366$ &$0.155$\\
    $E_{X_b}^n=25~\rm{MeV}$&$4\times 10^{-4}$ &$0.169$ &$0.092$ &$8\times 10^{-4}$ &$0.539$ &$0.178$\\
    $E_{X_b}^n=50~\rm{MeV}$&$3\times 10^{-4}$ &$0.143$ &$0.077$ &$5\times 10^{-4}$ &$0.518$ &$0.144$\\
    $E_{X_b}^n=100~\rm{MeV}$&$1\times 10^{-4}$ &$0.078$ &$0.041$ &$2\times 10^{-4}$ &$0.317$ &$0.075$\\ 
    \hline\hline
    \end{tabular}
    \label{Tab:Xb partial widths triangle diagrams sqrt ten mBstar}
\end{table}
\begin{figure}[htbp]
      \subfigure[] {
      \includegraphics[scale=0.438]{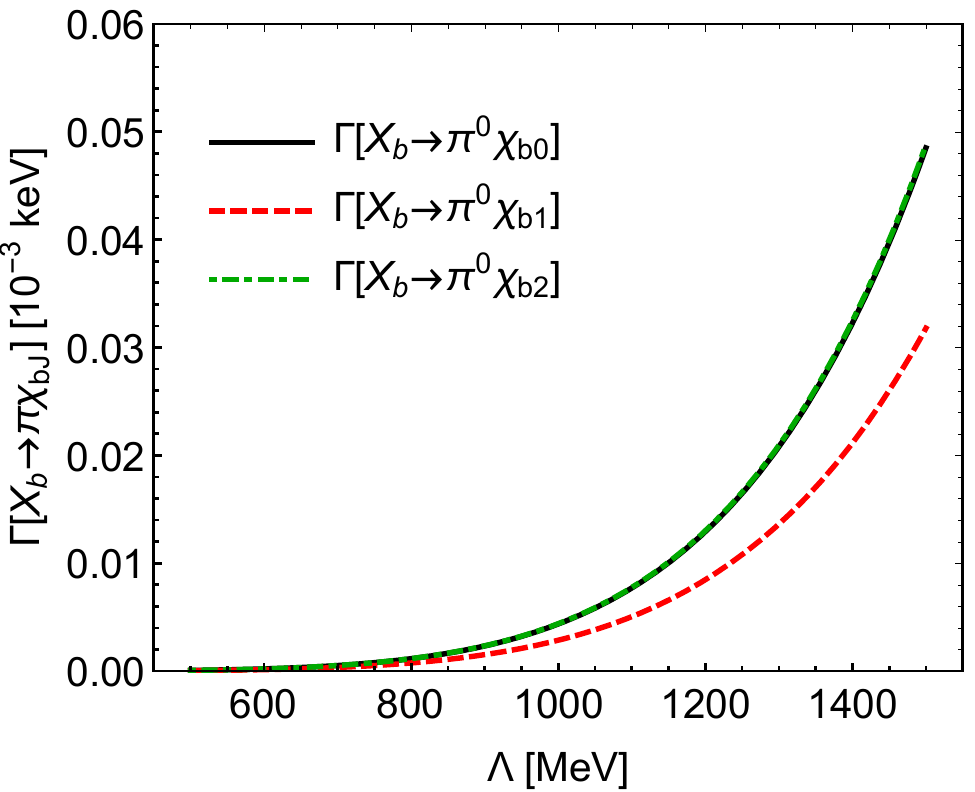}
      \label{fig_XbchibJpi0_mLambda}
     } 
      \subfigure[] {
      \includegraphics[scale=0.44]{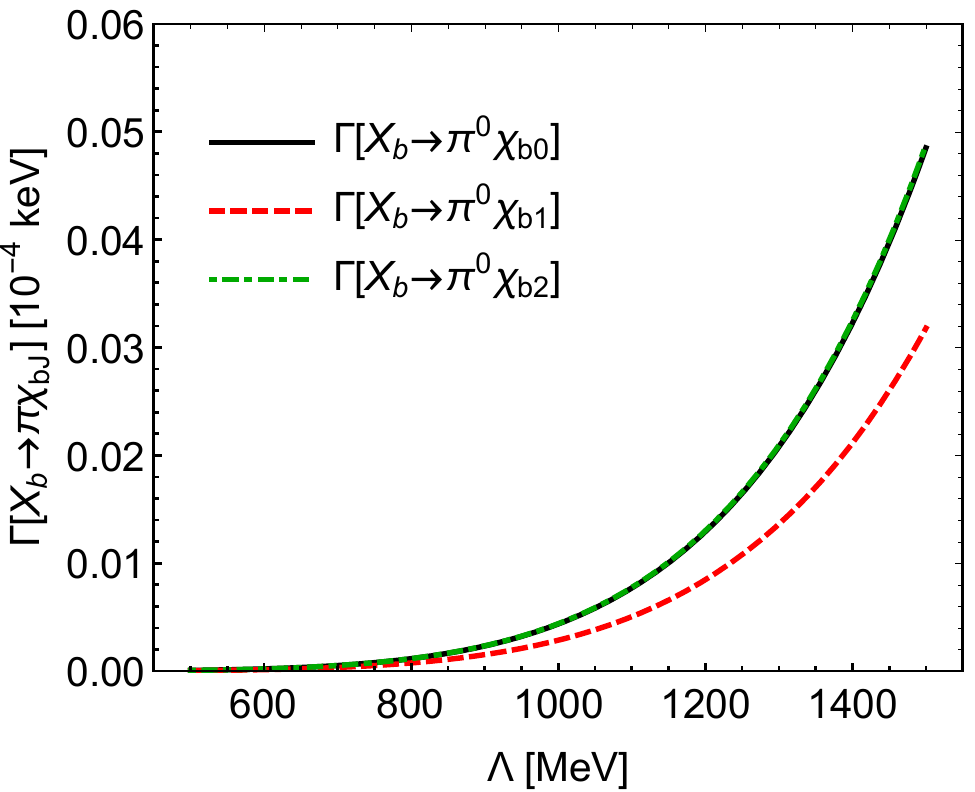}
      \label{fig_XbchibJpi0_sqrt10mLambda}
      }
    \caption{The contribution from the 2-point diagrams to the decay widths of $X_b \to \pi\chi_{bJ}$ ($J = 0, 1, 2$) with $\Lambda=0.5\sim 1.5$~GeV, $\varphi=0.813$, and $E_{X_b}^n=5$~MeV. The effective coupling $c_1$ is taken as $c_1 = g_1 / m_{B^*}$ in (a) and $c_1 = g_1 / (\sqrt{10} m_{B^*})$ in (b).}
    \label{fig_XbchibJpi0_Lambda}
\end{figure}
For the contributions of the bubble diagrams to the $X_b \to \pi \chi_{bJ}$ decay width, in our power counting scheme, the contributions from bubble diagrams are suppressed by $v^2$ compared with the triangle diagrams and therefore the bubble diagrams are not considered in the $X_b\to\pi\chi_{bJ}$ decay in our calculations in Sec.~\ref{sec:Numerical Results}. 

The 2-point integral $I[m_1, m_2]$ given in Appendix~\ref{sec:loop integrals} is regularization dependent, and the cut-off dependence of the two body decay widths from the bubble diagram contribution is shown in Fig.~\ref{fig_XbchibJpi0_Lambda} with $\Lambda=0.5\sim 1.5$~GeV, $\varphi=0.813$, and $E_{X_b}^n=5$~MeV. The coupling constant $c_1$ is also taken to be $g_1 / m_{B^*}$ and $g_1 / (\sqrt{10} m_{B^*})$ to  roughly estimate the contributions from the bubble diagrams, and the results are about $3\sim 4$ orders of magnitude smaller than the contributions of the triangle diagrams. The cutoff dependence of the decay widths is not significant for $\Lambda$ below $1$~GeV.


\bibliographystyle{apsrev}
\bibliography{Xb}
\end{document}